# Large-area femtosecond laser milling of silicon employing trench analysis

*Arun Bhaskar*[a,b], *Justine Philippe*[a], *Flavie Braud*[a], *Etienne Okada*[a], *Vanessa Avramovic*[a], *Jean-François Robillard*[a], *Cédric Durand*[b], *Daniel Gloria*[b], *Christophe Gaquière*[a], *Emmanuel Dubois*[a]*

[a]Univ. Lille, CNRS, Centrale Lille, Univ. Polytechnique Hauts-de-France, Yncréa Hauts-de-France, UMR 8520 - IEMN, F-59000 Lille, France

[b]STMicroelectronics, 850 Rue Jean Monnet, FR-38926 Crolles

## Abstract
A femtosecond laser is a powerful tool for micromachining of silicon. In this work, large-area laser ablation of crystalline silicon is comprehensively studied using a laser source of pulse width 300 fs at two wavelengths of 343 nm and 1030 nm. We develop a unique approach to gain insight into the laser milling process by means of detailed analysis of trenches. Laser scribed trenches and milled areas are characterized using optical profilometry to extract dimensional and roughness parameters with accuracy and repeatability. In a first step, multiple measures of the trench including the average depth, the volume of recast material, the average longitudinal profile roughness, the inner trench width and the volume removal rate are studied. This allows for delineation of ablation regimes and associated characteristics allowing to determine the impact of fluence and repetition rate on laser milling. In a second step, additional factors of debris formation and material redeposition that come into play during laser milling are further elucidated. These results are utilized for processing large-area (up to few mm$^2$) with milling depths up to 200 µm to enable the fabrication of cavities with low surface roughness at high removal rates of up to 6.9 µm$^3$ µs$^{-1}$. Finally, laser processing in combination with XeF$_2$ etching is applied on SOI-CMOS technology in the fabrication of radio-frequency (RF) functions standing on suspended membranes. Performance is considerably improved on different functions like RF switch (23 dB improvement in 2$^{nd}$ harmonic), inductors (near doubling of Q-factor) and LNA (noise figure improvement of 0.1 dB) demonstrating the applicability of milling to radio-frequency applications.

## 1. Introduction

Laser micromachining is a tool which is becoming increasingly relevant for micro/nanostructuring of materials. The use of ultrashort laser processing with pulse width in femtosecond range is advantageous because it can be applied on a wide range of materials like metals, semiconductors, dielectrics, alloys, and ceramics [1–4]. The range of surface processing is diverse varying from a small scale (a few nm$^2$) [5,6] to large scale (a few mm$^2$) [7–9]. Heat affected zone is greatly reduced for ultrashort lasers which allows for enhanced machining quality with low thermal impact of laser radiation on material.

Femtosecond micromachining of silicon has been reported previously in different studies. It has been explored for various applications like microfluidics [10], photovoltaics [11], IC characterization [7], hydrophobic surfaces [12], and laser dicing [13,14]. For this work, the targeted application is the fabrication of free standing membranes of SOI-CMOS RF circuits/functions on Silicon-on-Insulator (SOI) wafers. The term SOI-CMOS refers to CMOS processing technology used in semiconductor industry in combination with SOI wafer as the host substrate. SOI wafer comprises of a thick layer of handler silicon at the bottom, typically 750 µm, topped with a thin isolating buried oxide (BOX) layer and finally a thin layer of active silicon on the top. Microelectronic devices and circuits are realized into the topmost silicon active layer. The buried oxide is a thin layer of silicon dioxide which provides electrical isolation between active and handler silicon. Handler silicon is present for mechanical support and thermal dissipation. Despite the presence of an isolating BOX layer, handler substrate degrades electrical performance because it offers a parasitic coupling path to RF signals. By locally removing handler silicon under the active area of circuits, they can be suspended on the BOX in the form of a membrane. The motivation for creation of such membranes is to attain better RF performance with reduced loss and improved linearity as it has been reported previously in numerous similar studies [15–22].

The technique of laser milling is used in this work to create cavities underneath RF functions implemented in SOI-CMOS technology. The laser beam is raster scanned over the area to be milled. The starting thickness of SOI wafer is usually high (~750 µm) in order to prevent wafer warping during fabrication. As a preliminary step, the wafer is thinned down using processes like grinding and chemical mechanical polishing (CMP). The final thickness of silicon is much smaller and depends on the application requirements. With this in mind, the laser milling process is studied for depths up to 200 µm. The process is adaptable for geometries of side length ranging from ~100 µm to several mm by appropriately defining the laser milling trajectory.

The objective of this study is the development of a milling methodology that enables the control of surface quality and roughness while keeping the volume removal rate high. During micromachining of silicon, a commonly identified problem is the appearance of surface corrugations [13,23–27] which renders the surface cross section comb like as shown in Fig. 1. In the top view (Fig. 1c), these corrugations appear as microholes which can penetrate deep into the material depending on the process parameters used for milling. These corrugations are also observed in other materials like germanium [28]. They are not to be confused with the Laser Induced Periodic Surface Structures (LIPSS) which are ripples of the order of the wavelength of laser radiation [29,30]. In this study, this problem is tackled by choosing appropriate process parameters to minimize formation of corrugations in order to keep the roughness low while still having sufficiently high rates of removal.

In this work, a systematic approach is employed whereby step-by-step insight is gained towards best parameters for laser milling. In the first step, trenches (grooves) are scribed in silicon by linear displacement of the laser beam. Scribing allows quick estimation of impact of different process parameters on milling without performing two-dimensional surface machining. Section 3 reports different trench measures that are going to be studied and quantified by using optical profilometer measurements. The understanding of these parameters lays the foundation for high quality and high removal rate milling. Section 4 subsequently proposes the study of the surface laser milling to further understand the aspects of changing surface morphology, effect of debris, surface roughness and sidewall quality. The application of laser milling is presented in section 5 where membranes of RF circuits are fabricated. RF characterization of such circuits is performed in order to demonstrate the superior performance of the circuits on SOI membranes.

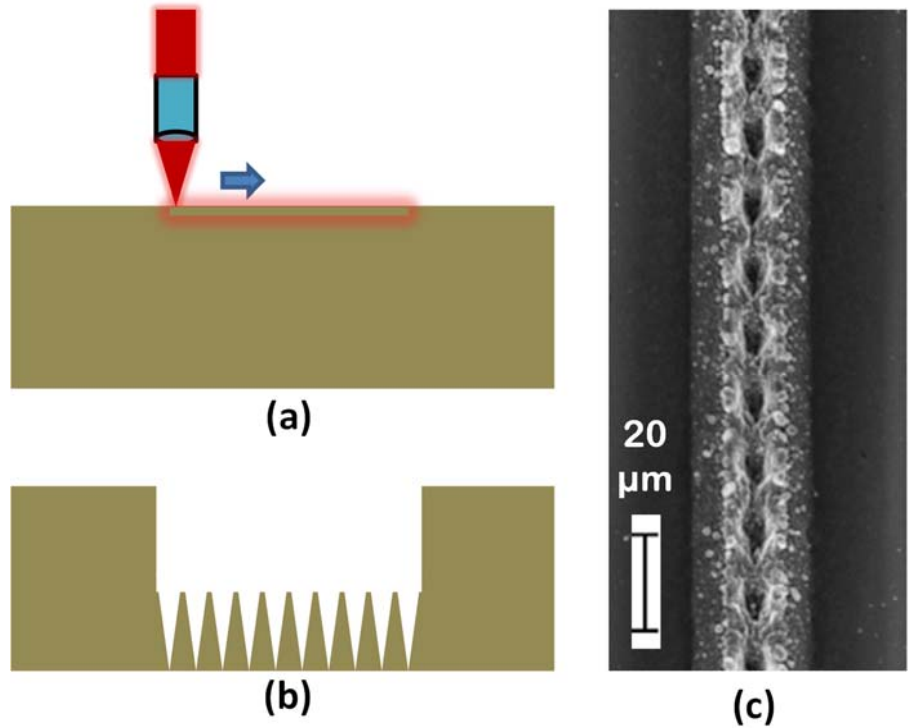

**Fig. 1.** (a) Linear translation of laser (b) Cross section of machined line and (c) Top view of machined line

## 2. Experimental description

### 2.1 Laser system

The experiments are performed under ambient conditions using a Diode Pumped Solid State (DPSS) ultrafast fiber laser from Amplitude Systèmes (Tangerine) with a fundamental output at 1030 nm and pulse width (FWHM) of 300 fs. Additional wavelength output of 343 nm is generated using a harmonic box (Amplitude-Systèmes). The 343 nm wavelength is obtained by frequency tripling realized by cascading non-linear crystals, beginning with frequency doubling of the fundamental input beam at 1030 nm and subsequent sum frequency generation of both waves. The maximum power output available at the workpiece is 1.2 W and 13 W for 343 nm and 1030 nm sources, respectively. Power attenuation is achieved by cascading a rotating birefringent half-wave plate with a polarizer which provides a continuous variation of the laser power while decoupling the power attenuation from the laser repetition rate. The pulse repetition rate for maximum efficiency of laser output is 200 kHz and can go up to 2 MHz with reduced efficiency. The laser beam displacement is effected by a galvanometric scanner (Canon) followed by a flat-field telecentric lens (Sill Optics) with a focal length of 100 mm. The maximum galvo speed for milling without path distortions is 20 mm s$^{-1}$ and 100 mm s$^{-1}$ for 1030 nm source and 343 nm source respectively [15]. The beam waist (radius) is determined experimentally using the regression of squared diameter on fluence as first proposed by Liu et al. [31]. For 343 nm and 1030 nm sources, the measured beam waist on silicon is 5.8 ± 0.3 µm and 8.2 ± 0.2 µm, respectively. A quarter wave ($\lambda/4$) plate is placed before the lens to obtain a circular polarization of the laser beam.

Laser micromachining is performed on High-Resistivity (HR) SOI substrate with resistivity > 1 kΩ.cm. This value of resistivity is 2-3 orders of magnitude higher than normal silicon wafer. However, the interpretations in this study is applicable for any type of silicon as it has been shown that substrate resistivity does not affect the ablation threshold for multi-pulse ablation [32]. The laser machined lines and cavities are characterized using optical profilometry, optical microscopy and scanning electron microscopy (SEM). The optical profilometer (Bruker ContourGT™-X) is a non-contact profiler based on vertical scanning interferometry using a broadband white light source. The SEM (Zeiss Supra55VP) scans are taken at an acceleration voltage of 10 kV.

For SOI-CMOS application demonstration, gas phase $XeF_2$ etching is used to selectively remove silicon with respect to silicon dioxide. An etch-protect layer (Ajinomoto GX-T31) is laminated on the back side of the SOI die. This ensures that silicon is removed completely only in the areas where cavities are milled leaving the BOX layer intact. The etching is performed in a Xactix $XeF_2$ etcher.

## 2.2 Choice of parameters for trench scribing

Scribing of trenches is the first stage of our analysis to study the different ablation regimes and to determine the fluence and repetition rate suitable for milling application. The parameters that are varied are the pulse repetition rate ($f_{rep}$), pulse energy ($E_p$) and scan speed ($v_{scan}$). From the chosen parameters, the average power can be given by the expression:

$$P_{avg} = E_p f_{rep} \quad (1)$$

To obtain high removal rates, a maximum average power of >1 W up to ~2.5 W is used for removal of trenches. Although higher power is supported in the laser system for the 1030 nm line, it is not utilized because surface roughness and volume removal rate are degraded at high power as it shall be discussed in section 3.2. In order to obtain a given average power, the pulse repetition rate and the pulse energy can be independently chosen. When the pulse energy is varied, correspondingly the peak fluence of the laser pulse changes as given by the expression:

$$\phi_0 = \frac{2E_p}{\pi w_0^2} \quad (2)$$

Here, $w_0$ is the beam waist radius, the radial distance measured from beam cross section at focal plane where the intensity reduces to $1/e^2$ times the value at the centre. All fluence values reported hereinafter corresponds to the peak fluence. Three cases of trench scribing have been studied as shown in Table 1. For the first two cases, 200 kHz repetition rate is used as this is the repetition rate at which laser efficiency peaks. For the third case repetition rate is reduced to 30 kHz so that power corresponding to maximum pulse energy is sufficiently high (~1.8 W). While these cases have been briefly treated in [16], this paper outlines a detailed analysis into each case. Three scan speeds are studied per case. Each scribed trench is characterized using optical profilometry. The different analyses that are performed for each scribed trench are: average trench depth, recast layer volume, average profile roughness, average trench width and volume removal rate. A graphical illustration of measured parameters can be found in Appendix A.1. Each analysis is presented separately in section 3.

**Table 1.** Different cases in trench scribing study

| Case | Repetition rate (kHz) | Average Power (W) | Fluence ( J cm$^{-2}$) | Wavelength (nm) |
|---|---|---|---|---|
| 1 | 200 | 0.08 – 1.21 | 0.76 – 11.67 | 343 |
| 2 | 200 | 0.18 – 2.56 | 0.86 – 12.11 | 1030 |
| 3 | 30 | 0.09 – 1.79 | 2.59 – 56.5 | 1030 |

The uncertainty on measured depths is less than 1% owing to the high accuracy of vertical scanning interferometry. The accuracy of volume measurements is primarily governed by the evaluation of distances in the plane of the sample surface. This accuracy is a function of the spatial sampling determined by the magnification of the lens and the pixel capacity of the observation camera.

In this study, we estimate that lateral distances are affected by a maximum fixed error of 1 µm corresponding to 2 spatial sampling pixels. The power meter accuracy is less than 3%. Based on these figures, the uncertainty in fluence is estimated as < 13%.

## 3. Scribing of trenches

This section presents the discussion regarding the different experimental observations made in the scribing of trenches utilizing the parameters detailed in section 2.2.

### 3.1 Ablation regimes in micromachining of silicon

Laser micromachining is essentially multi-pulse ablation wherein different ablation regimes can be identified. Each ablation regime is characterized by a certain dominant physical mechanism of ablation and the regimes are separated by a threshold fluence, which changes as a function of the process parameters that are employed. In this section, the average trench depth and recast layer volume are studied to distinguish the ablation regimes.

### 3.1.1 Average trench depth ($d_t$)

The optical profilometer renders a 3D profile of the trench. A 1D height profile is taken approximately along the central axis of the trench. Average trench depth is determined by summing the different height values comprising the axial trench profile and dividing by the number of points as given by the equation:

$$d_t = \frac{1}{N}\sum_1^N z_i \tag{3}$$

For the 3 cases, the plots of average trench depth as a function of fluence is shown in Fig. 2. The depth plotted on a semi-logarithmic scale shows two ablation regimes with different ablation thresholds. This is a well-known phenomenon which has been reported in metals [1,33], semiconductors [23,34] and dielectrics [35,36]. The depths in the two regimes can be given by

$$d_t^{g,s} = l_{g,s}\, ln\left(\frac{\phi_0}{\phi_{th}^{g,s}}\right) \tag{4}$$

where $d_t^{g,s}$ is the average trench depth, $\phi_{th}^{g,s}$ the threshold fluence and $l_{g,s}$ the characteristic ablation depth with the subscript/superscript index 'g' and 's' standing for gentle and strong ablation regimes. The transition fluence between these two regimes is determined by the scan speed when all other laser parameters are fixed. Higher scan speeds allow utilization of higher fluence (consequently a higher average power) in the gentle ablation regime. Thus, higher removal rates can be obtained by using higher scan speeds while still operating in the gentle ablation regime. The ablation regime is important as it determines several aspects like recast layer volume, roughness, trench width narrowing and a detailed discussion of this is presented in sections 3.3 – 3.5.

From the plots of average depth as a function of fluence, curve fitting is performed to associate data points with the two regimes. The extracted values of characteristic depth and threshold fluence for the 3 cases at different speeds are tabulated in Table 2. Sample trench profiles for the two regimes have been provided in appendix A.2. With increase in wavelength, both the characteristic depths ($l_{g,s}$) and threshold fluences ($\phi_{th}^{g,s}$) increase at a given speed. In the gentle ablation regime, at repetition rate of 200 kHz, the difference in ablation thresholds between 343 nm and 1030 nm source is small. For the 1030 nm source at a repetition rate of 200 kHz, a good point of comparison with nearly the same laser system as ours is provided in the work of Shaheen et al [37]. They reported a threshold fluence of 0.54 and 2.4 J cm$^{-2}$ in the two regimes for 20 pulses. This correlates excellently with values obtained for scan speed of 50 mm s$^{-1}$ where pulse overlap is small (0.25 µm) . For the 1030 nm source, at a repetition rate of 30 kHz, the threshold fluence for gentle ablation is much larger as compared to other two cases. The characteristic depths in the gentle ablation regime are smaller at same speed for this case. Higher threshold fluence and lower characteristic depths do not necessarily mean that ablation rate is reduced. This is because the width of the ablated trench also needs to be considered to calculate the volume removal rate. The width of trench is dependent on the experimental parameters and is discussed in section 3.5. The material removal rates for the 3 cases are presented in section 3.6.

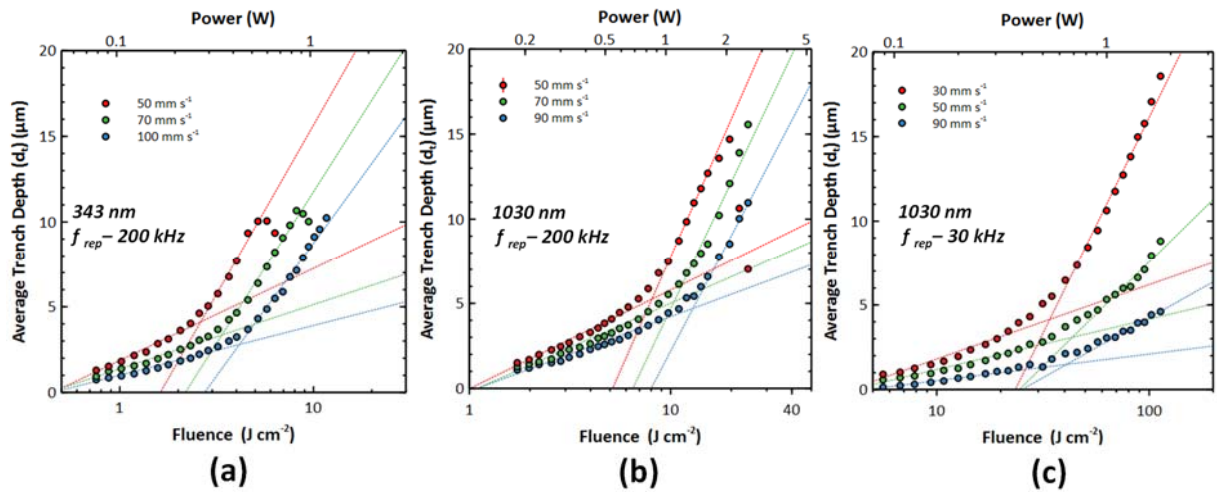

**Fig. 2.** Average trench depth as a function of fluence for 3 cases at different scan speeds (a) source – 343 nm, $f_{rep}$ – 200 kHz (b) source – 1030 nm, $f_{rep}$ – 200 kHz and (c) source – 1030 nm, $f_{rep}$ – 30 kHz. The dashed lines represent the fitting of points into two different regimes with x-intercept being the threshold fluence

For the 200 kHz repetition rate cases, average depth measured from axial trench profile starts to decrease at a high fluence. This is observable at speeds of 50 and 70 mm s$^{-1}$ for 343 nm source and at 50 mm s$^{-1}$ for 1030 nm source. This is mostly likely due the formation of debris at the bottom of the trench as observed in [23] and supported by explanation given in Appendix A.3.

**Table 2.** Characteristic depth and corresponding threshold fluence in the two regimes for 3 sets of process parameters (repetition rate and wavelength) at 3 scan speeds. The final column contains threshold fluence extracted from recast layer volume in the strong ablation regime

| Parameter | Speed (mm s$^{-1}$) | $l_g$ ($\mu m$) | $\phi_{th}^g$ ($J\,cm^{-2}$) | $l_s$ ($\mu m$) | $\phi_{th}^s$ ($J\,cm^{-2}$) | $\phi_{th}^{Vs}$ ($J\,cm^{-2}$) |
|---|---|---|---|---|---|---|
| λ=343 nm $f_{rep}$=200 kHz | 50 | 2.34 | 0.45 | 8.60 | 1.62 | 1.86 |
|  | 70 | 1.63 | 0.43 | 7.72 | 2.18 | 2.88 |
|  | 100 | 1.26 | 0.45 | 6.75 | 2.74 | 3.61 |
| λ=1030 nm $f_{rep}$=200 kHz | 50 | 2.53 | 0.5 | 11.71 | 2.58 | 2.12 |
|  | 70 | 2.26 | 0.55 | 10.81 | 3.24 | 4.11 |
|  | 90 | 1.91 | 0.55 | 9.83 | 4 | 5.26 |
| λ=1030 nm $f_{rep}$=30 kHz | 30 | 1.91 | 1.94 | 11.12 | 11.73 | 6.08 |
|  | 50 | 1.29 | 2 | 5.38 | 12.25 | 5.94 |
|  | 90 | 0.68 | 2.34 | 3.07 | 12.6 | 2.96 |

The higher ablation threshold seen in the gentle ablation regime for 1030 nm source at a repetition rate of 30 kHz is attributed to the higher fluence used. It has been reported that for fluence > 4 Jcm$^{-2}$, the ablation enters a different regime and is dominated by hydrodynamic motion of silicon [38]. In all cases the increasing threshold fluence at higher speed is because of the fact that effective number of pulses hitting a spot reduces at higher speed. In the work of Ashkenasi et al., it was shown that when number of pulses are high, the transition from gentle to strong regime happens at a lower fluence and this explains the threshold fluence trend with respect to speed [35].

In the detailed study of grooves (trenches) in silicon by Crawford et al., it is argued that scribing of trenches may not be equivalent to stationary multi-pulse ablation [23]. For scribing of trenches, there is a spatial displacement of laser beam with respect to the sample from one pulse to the next. Each pulse causes ablation and modification of the surface. Thus, for subsequent pulses, the laser energy is incident on an uneven and changing surface which gives rise to local variation of deposited energy. While this is also true for stationary ablation, the difference is that geometry of the surface encountered during ablation involving translation is much more skewed making the local energy distribution more uneven. Additionally, they found that at low translation speeds, a nearly linear dependence of trench depths as a function of pulse energy. At the same time, on a logarithmic scale, they could fit the data points into two regimes. The authors preferred to use the two regimes model to interpret their data. However, they expressed the view that linear dependence of ablation depths cannot be ruled out and that two different regimes may not be present.

In this study, the same question had to be addressed: is a linear or two-regime model better to explain ablation data? In order to answer this question, several analyses of the recast layer are performed. It will become clear in next subsection that two ablation regimes are indeed present for all 3 cases.

3.1.2 Recast layer volume ($V_r$)

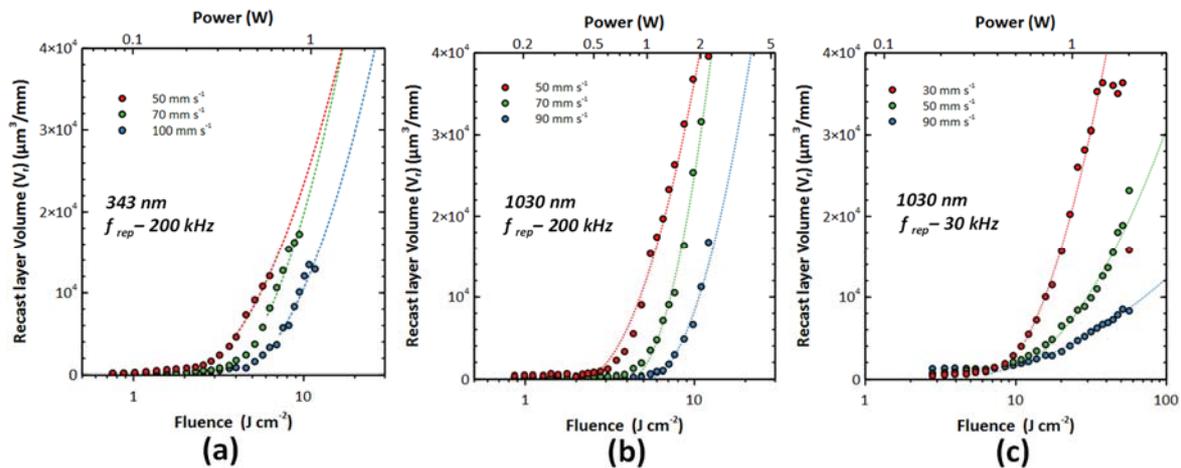

**Fig. 3.** Recast layer volume as a function of fluence for 3 cases at different scan speeds (a) source – 343 nm, f$_{rep}$ – 200 kHz (b) source – 1030 nm, f$_{rep}$ – 200 kHz and (c) source – 1030 nm, f$_{rep}$ – 30 kHz. . The dashed lines represent the fitting of data points in strong ablation region with x-intercept being the threshold fluence

Recast layer is defined as the aggregate of debris/redeposited material in the vicinity of the ablated trench. In the profilometer scan, the points representing the recast layer form a 2D surface extending above the substrate. The recast layer volume is calculated as the volume enclosed by this surface and the reference level of the substrate (zero level). The recast layer volume obtained for the three cases is shown in Fig. 3. The unit µm³ mm$^{-1}$ refers to the volume of recast layer in µm³ when scribing a trench of length 1 mm. This measure is a key indicator of the presence of two ablation regimes. For all 3 cases, the volume of recast layer is low in the gentle ablation regime followed by an abrupt increase which shows the onset of strong ablation. At a given speed, there is only a very small increase in ablation volume when going from lowest to highest fluence within the gentle ablation regime. Higher fluences out of this regime result in strong ablation and this is accompanied by an increase of volume of recast layer. In the strong ablation regime, the volume of recast layer is much more sensitive to the fluence. A large difference in volume is observed between the highest and lowest fluence conditions within the strong ablation regime at a given speed. For the 1030 nm source and repetition rate of 30 kHz, at speed of 30 mm s$^{-1}$, the volume of recast layer starts to decrease at very high fluences. When the profilometer scan is observed, it is seen that some parts of the recast layer is missing along the sides of the trench. This can be attributed to strong expulsion of ablated material at such fluences which possibly impedes the formation of recast layer at certain places along the sides of the trench.

A threshold fluence is also extracted using the recast layer volume data. An assumption is made that the recast layer volume ($V_r$) is directly proportional to the ablated volume of the silicon. The threshold fluence for gentle ablation regime has not been reported because the measured volume is close to the lower limit of the profilometer measurement of volume and hence it is difficult to obtain a reliable curve fit. The ablated volume of silicon follows a squared logarithm dependence with respect to the fluence [39]. Accordingly, the equation for recast layer volume data can be given by:

$$V_r = k . ln^2 \left(\frac{\phi_0}{\phi_{th}}\right) \qquad (5)$$

Least squares curve fit has been performed and the threshold fluence in strong ablation regime is depicted as $\phi_{th}^{Vs}$ in Table 2. It can be seen that the threshold fluence for 200 kHz repetition rate for both cases are slightly higher than those obtained by trench scribing. However, it is vastly different for 1030 nm source with repetition rate of 30 kHz. The reason could be that the assumption of recast layer volume being proportional to ablated volume of silicon is not valid at high fluences.

## 3.2 Impact of ablation regime on micromachining

In the previous section, ablation regimes have been studied. This section outlines how the ablation regime affects the micromachining in terms of three important parameters: average roughness, trench width and volume removal rate.

### 3.2.1 Average roughness of trench profile ($R_a$)

One of the key aims in this study is to determine the fluence and repetition rate which minimize corrugations. Trench profiles have been presented and corrugations have been studied for different experimental conditions in our previous work in [16]. Here, corrugations have been quantified by using average roughness of axial trench profile. Higher roughness gives trench morphology with microholes that go deeper into the material. The expression for average roughness can be given by:

$$R_a = \frac{1}{N} \sum_1^N |z_i - d_t| \qquad (6)$$

The average roughness measured from the axial trench profile using equation 6 is plotted in Fig. 4. It can be seen that for the two 200 kHz repetition rate cases, very high roughness values of 2 – 3 μm can be reached. The profile is highly corrugated at high fluences. For the two cases, the increase in trench profile roughness shows the same behaviour as the recast layer volume. A nearly constant roughness is seen in the gentle ablation regime followed by an increase in roughness with respect to fluence in the strong ablation regime. Interestingly, for the 1030 source at a repetition rate of 30 kHz, the surface roughness shows a different behaviour. It is nearly constant over both regimes with only a small increase at high fluences and the maximum value is < 1 μm.

The formation of a periodic structure of holes with a spatial separation well above the laser firing repetition pitch has been reported in the literature. This effect has been observed on silicon where the holes thus formed are called craters or cones [24,40] .The formation of such holes has been used for the fabrication of superhydrophobic surfaces in the work of Yong et al. [41]. They attribute the appearance of holes to the dynamic accumulation of recast material along the machining path. The hypothesis is that initial few laser pulses cause slow build-up of recast material ahead of the current laser position. In this region, depth of removal of silicon is high. For the next few pulses, the laser spot encounters the recast material projecting over the silicon surface. In this region, overall depth of removal of silicon remains low because part of energy is lost to removal of accumulated recast material. This cycle repeats and results in the formation of a periodic structure.  We find that this hypothesis is applicable in our experiments as well and there are two supporting arguments. For high repetition rate and low fluence cases (Fig. 4a and 4b), the roughness increases abruptly in a fluence regime where recast layer volume also increases abruptly (Fig. 3a and 3b). Additionally, the inner trench width also reduces for these two cases suggesting a build-up of recast material. For the case with low repetition rate and higher fluences, the roughness (Fig. 3c) does not change abruptly even when the recast layer volume does. This can be due to two reasons. The first is that when the repetition rate is 30 kHz, the spacing between pulses is one order of magnitude higher as compared to 200 kHz which could prevent localized accumulation of recast silicon. The second is that higher fluence results in more vigorous expulsion of ablated material [42], which again possibly prevents local accumulation of recast material .

The periodic nature of the profile can be easily noticed in the SEM scans depicted for a limited set of processing parameters in Appendix A.4. It is found that the period of the profile is proportional to the scan speed. For the 343 nm source, a Fourier analysis of profiles revealed periods of 6.9, 9.6, and 13.8 µm for speeds of 50, 70 and 100 mm s$^{-1}$, respectively. The periodicity is found not to depend on the fluence. However, the amplitude of the dominant period increases with fluence.  While average roughness is one measure of roughness that has been presented, Root Mean Squared (RMS) roughness may also be used. Same trends can be observed by plotting the RMS roughness as a function of fluence and is not separately shown here.

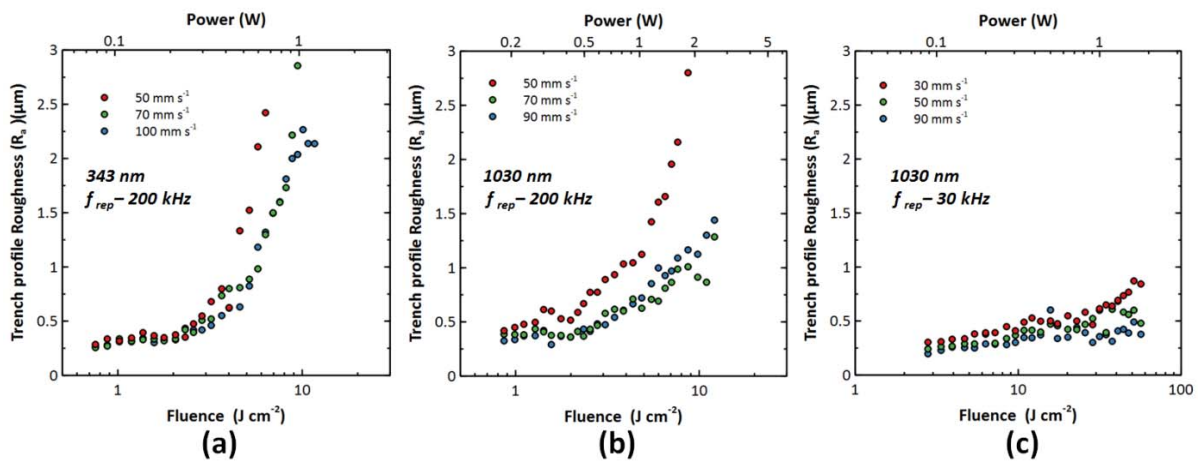

**Fig. 4.** Average roughness of trench as a function of fluence for 3 cases at different scan speeds (a) source – 343 nm, $f_{rep}$ – 200 kHz (b) source – 1030 nm, $f_{rep}$ – 200 kHz and (c) source – 1030 nm, $f_{rep}$ – 30 kHz.

### 3.2.2 Trench width ($W_t$)

For the measurement of trench width, the 1D cross-sectional profiles are taken perpendicular to the axis of the trench. The width of the recast layer can vary considerably along the length of the trench. Hence, in order to obtain an average width, 10 cross-sectional profiles are taken at different points along the length of the axis. The average widths calculated from these profiles are plotted in Fig. 5. Only the inner trench widths are shown. When the outer trench widths were plotted, a large scatter is observed and no clear trend can be seen. Hence, these data are not presented here.

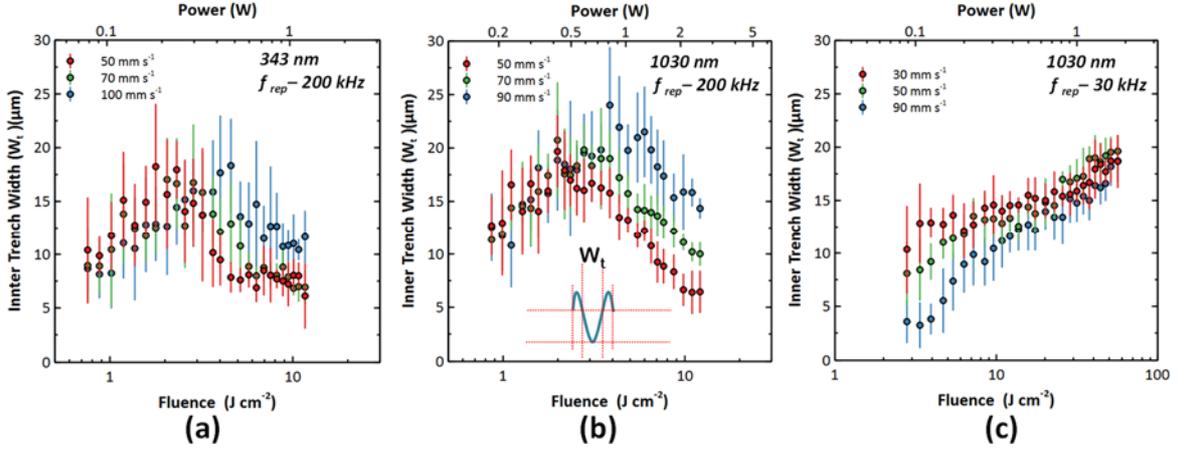

Fig. 5. Inner trench width as a function of fluence for 3 cases at different scan speeds (a) source – 343 nm, $f_{rep}$ – 200 kHz (b) source – 1030 nm, $f_{rep}$ – 200 kHz (c) source – 1030 nm, $f_{rep}$ – 30 kHz. The error bars represent symmetric error of one standard deviation.

The 343 nm source features a smaller beam waist of as compared to 1030 nm. The width of cut scales directly with the beam waist as given by the equation:

$$W_t = \sqrt{2w_0^2 \ln\left(\frac{\phi_0}{\phi_{th}}\right)} \qquad (7)$$

Hence a bigger width of cut can be expected for 1030 nm source. However, it is also determined additionally by the threshold fluence which is smaller for 343 nm source. On comparing the two 200 kHz repetition rate cases, an overall observation indicates higher width of cut is indeed seen for 1030 nm source. However, the scatter of data is not negligible for these data points, which makes it difficult to compare the two cases. Nevertheless, it is seen for both cases that in the gentle ablation regime, the width of trench increases. An additional important observation is made for the two cases. With the onset of strong ablation regime, the trench width starts to decrease. This is most likely due to accumulation of recast material on the sidewalls of the trench as reported in [23]. The reduction in trench width makes the milling process less efficient. This is because for each subsequent line traversed in the milling trajectory, a considerable amount of energy is deposited on the recast layer which does not result in any net ablation from the substrate. For the last case of the 1030 nm source at a repetition rate of 30 kHz, the trench widths in the gentle ablation regime is more sensitive to the fluence especially at high scanner speeds. A very low width of cut of ~3 μm can be obtained at a scanner speed of 90 mm s$^{-1}$. The scatter of data is smaller as compared to the other two cases indicating the presence of a more uniform recast layer. The narrowing of trench width in the strong ablation regime is not seen for this case. This would be useful in preventing loss of laser energy to recast layer and improving milling efficiency.

### 3.2.3 Volume removal rate ($VRR$)

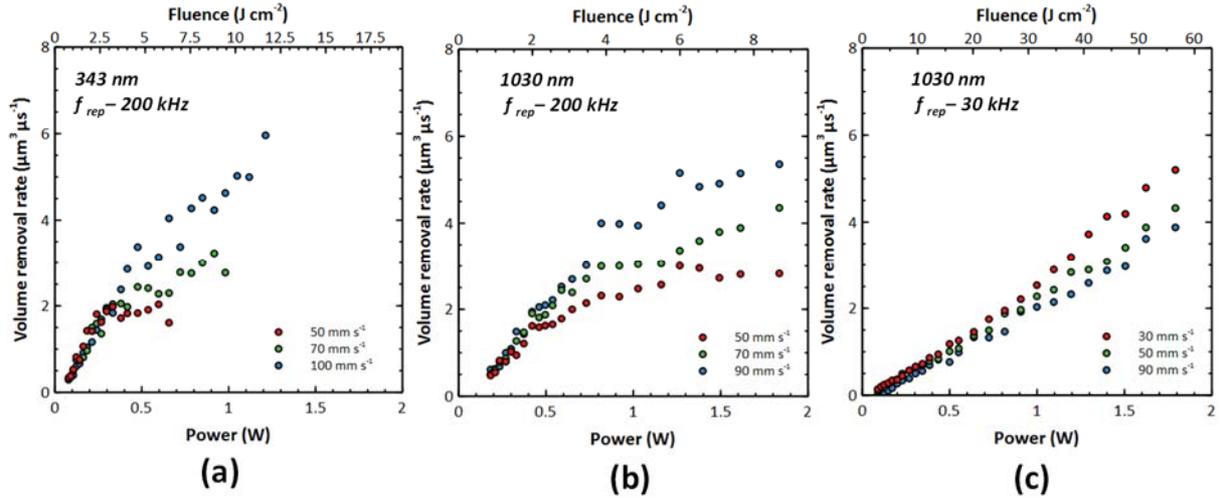

**Fig. 6.** Volume removal rate as a function of average power for 3 cases at different scan speeds (a) source – 343 nm, $f_{rep}$ – 200 kHz (b) source – 1030 nm, $f_{rep}$ – 200 kHz (c) source – 1030 nm, $f_{rep}$ – 30 kHz.

The volume removal rate is an important factor which determines how fast material can be removed from the substrate. It is computed based on the empirical observation that the trench cross section can be well approximated to be a triangle using the formula:

$$VRR = \frac{d_t W_t v_{scan}}{2} \qquad (8)$$

The volume removal rates are compared as a function of average power and the plots for the 3 cases are shown in Fig. 6. An overall comparison in Fig. 6 shows that the high repetition rate cases (a) and (b) have higher material removal rates as compared to case (c) within the range of experimental parameters chosen. This difference is more distinct by observing the slopes at low laser power. For cases (a) and (b), with increasing power, the removal rate tends to flatten out. This is because of formation of debris which reduces both trench width and depth. The maximum value is higher for higher scan speeds. This phenomenon is not seen for case (c) and the removal rate steadily rises even at high power. The trench narrowing effect and corresponding reduction of VRR for cases (a) and (b) can be potentially reduced by using a larger beam waist which gives a wider cut and helps easier ejection of debris. This also provides an added advantage of even higher removal rate because threshold fluence reduces with increasing beam waist [43]. However, this compromises the resolution of the laser milling which is more important in our application.

In summary, laser scribing of trenches has been studied for 3 different cases leading to the following important observations: i) the use of the 1030 nm source at a lower repetition rate and higher fluence gives a lower material removal rate, ii) however, other advantages, such as reduction of debris, low profile roughness for deep trenches, absence of trench narrowing effect, make this case suitable for usage in laser milling. Thus, the multiple analyses based on i) the ablation average depth, ii) the volume of recast material, iii) the average longitudinal profile roughness, iv) the inner trench width and v) the volume removal rate made it possible to identify the ablation regimes and to determine the impact of the fluence and repetition rate on these parameters. Additional factors come into play during laser milling. This is discussed in the following section.

## 4. Milling of cavities

Section 3 provided an overview of trench scribing with different parameters and the optimal values have been highlighted. In this section, further analysis is presented for milling over a surface area using the chosen parameters and how process parameters are further refined in order to avoid high surface roughness.

Laser milling is performed by raster scanning the laser beam with line to line spacing (pitch) of 10 µm. Square cavities are milled with side length ranging from 100 µm to 1 mm. The parameters of interest are surface morphology, roughness and quality of sidewall. The repetition rate of 30 kHz is chosen to avoid microhole formation. The scan speed is set to 20 mm s$^{-1}$ and the fluence is varied. The first study is made by a single scan over the milling area. A trench is also scribed using the same parameters to make a comparison between 1D machining and 2D milling. The SEM scans of the surface of the cavities for one scan of laser beam are shown in Fig. 7. It can be seen that overall morphology looks good with presence of few microholes notably at a fluence of 4.7 J cm$^{-2}$.

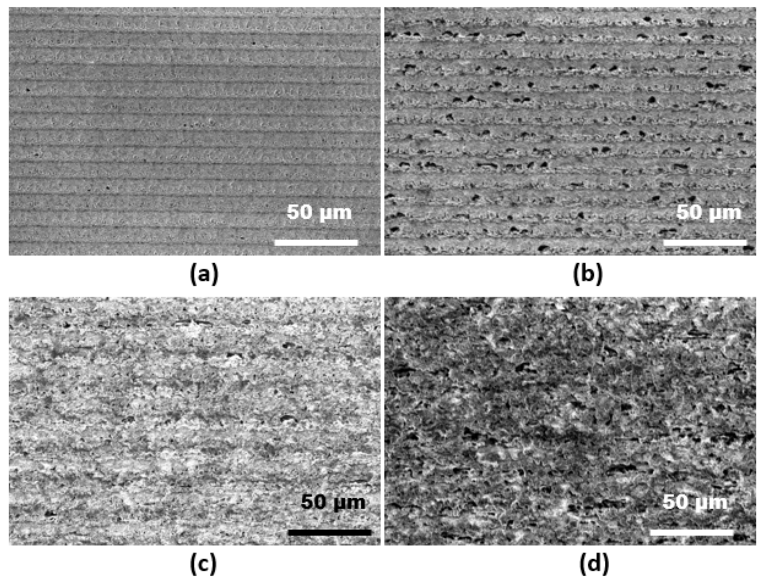

**Fig. 7.** SEM image of cavity surface after ultrasonic cleaning. Milling parameters: 1030 nm; $N_{scans}$ = 1; Scan speed - 20 mm s$^{-1}$; Repetition rate - 30 kHz; Pitch - 10 µm; Fluence (a) 2.8 J cm$^{-2}$ (b) 4.7 J cm$^{-2}$ (c) 12.7 J cm$^{-2}$ (d) 17.4 J cm$^{-2}$

**Table 3.** Calculated quantities for milled cavities (single scan) on 1030 nm source at repetition rate of 30 kHz and scan speed of 20 mm s$^{-1}$

|  | Fluence (J cm$^{-2}$) | | | |
| --- | --- | --- | --- | --- |
| **Parameter** | **2.8** | **4.7** | **12.7** | **17.4** |
| **Before ultrasonic cleaning** | | | | |
| Average depth of cavity (µm) | 1.3 | 1.5 | 2.3 | 4.1 |
| **After ultrasonic cleaning (in IPA for 5 minutes)** | | | | |
| Average depth of cavity (µm) | 1.9 | 2.9 | 6 | 8.9 |
| Average depth of trench profile (µm) | 1.9 | 4.2 | 6.9 | 10.4 |
| Average roughness of trench profile (µm) | 0.2 | 0.4 | 0.4 | 0.6 |
| Average roughness of cavity (µm) | 0.2 | 0.5 | 0.8 | 1.2 |
| RMS roughness of trench profile (µm) | 0.2 | 0.5 | 0.6 | 0.7 |
| RMS roughness of cavity (µm) | 0.2 | 0.7 | 1 | 1.5 |
| Volume removal rate of cavity (µm$^{-3}$ s$^{-1}$) | 0.36 | 0.55 | 1.14 | 1.69 |

Several parameters are computed for single scan milling from the profilometer scans of the milled cavity and the scribed trench which is summarized in Table 3. It can be seen that depth of cavity after ultrasonic cleaning increases in all cases. With increase in fluence, the difference is larger. This means that a significant quantity of material at the bottom of the cavity is loosely bound suggesting a redeposition of ablated material. At the lowest fluence of 2.8 J cm$^{-2}$, the roughness and depth values for the cavity and trench are excellently correlated. With increasing fluence, the difference in depth and roughness parameters between cavity and trench profile becomes larger. This increase in roughness can be attributed to presence of the ablation plume and redeposited material in the vicinity of the ablation area which interacts with the laser beam [44,45]. Ablation plume is the collection of atoms, clusters, ions and electrons that is emitted during ablation which can dynamically interact with the laser beam during the milling process. For the chosen set of parameters, high removal rates are obtained in the range of $10^5 – 10^6$ μm$^3$s$^{-1}$.

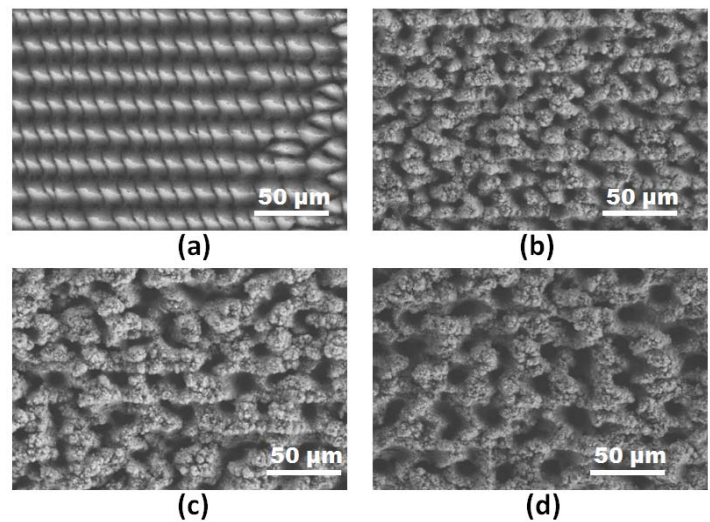

**Fig. 8.** SEM image of the bottom of the trench after cleaning in ultrasonic bath as seen on SEM. Multipass milling (Depth ~ 100 μm) using 1030 nm source with parameters: Scan speed - 20 mm s$^{-1}$; Repetition rate - 30 kHz; Pitch - 10 μm; Fluence/ Number of scans (a) 2.8 J cm$^{-2}$/65 (b) 4.7 J cm$^{-2}$/28 (c) 12.7 J cm$^{-2}$/17 (d) 17.4 J cm$^{-2}$/10

Having studied milling for a single pass, the number of passes is increased to mill cavities with a depth of ~100 μm. The morphology of the surface of the cavity is shown in Fig. 8. It is drastically different from that observed for single pass. A large number of microholes are seen on the cavity surface. The number of holes decreases while the size of the holes increases at a higher fluence. At the lowest fluence of 2.8 J cm$^{-2}$, a regular array of microstructures approximately in the form of pyramids is seen. The appearance of such self-organized microstructures has been reported for different metals like Cu, Al, Ti etc. [46]. Same has been observed for silicon in numerous studies [11,42,47]. The reason for formation of such ordered structures is formation of regular array of bumps in the initial few passes of laser in the milled area. This gradually evolves into a surface morphology containing pillars due to local concentration of laser energy in the valleys and accumulation of ablated matter on the top of the pillars. A visual depiction of this phenomenon can be observed in the work of Lee et al. where SEM images for different number of passes has been taken [42]. They also found that the ordered microstructures disappear at higher fluence which is consistent with this work. The reason has been attributed to strong expulsion of ablated matter which prevents formation of ordered structures. Additionally, it is observed that for the low fluence of 2.8 J cm$^{-2}$, a lot of debris are formed and the smaller cavities are almost fully covered in particle debris. For higher fluences, the quantity of debris is much lower and mostly found in the regions surrounding the ablated cavity.

The average depth is plotted in Fig. 9a for all cavities before and after ultrasonic cleaning to quantify the amount of redeposited material. The x-axis in this graph represents the side length of square cavity being milled. It can be seen that in all cases redeposited matter exists at the bottom of the cavity for all cases. For fluences of 2.8 J cm$^{-2}$ and 4.7 J cm$^{-2}$, milled depth is less for smaller cavities. The average difference in depths before and after ultrasonic cleaning are 3, 1.7 and 1.5 µm for fluences of 4.7, 12.7, and 17.4 J cm$^{-2}$ respectively. For fluence of 2.8 J cm$^{-2}$, the data before ultrasonic cleaning is not taken because many cavities are covered fully in particle debris. In Fig. 9b, the average depths are found in a similar way but the number of passes is doubled to have nearly twice the depth. The average difference in depths before and after ultrasonic cleaning are 5.2 and 3.7 µm for fluences of 12.7 and 17.4 J cm$^{-2}$, respectively. Thus, when the cavity becomes deeper, the quantity of redeposited material increases. An efficiency factor is determined using the following formula which is one way to quantify how efficiently laser energy is utilized in the milling process:

$$\eta_f = \frac{d_{cav}}{N_p d_t} \quad (9)$$

where $d_{cav}$ is the average depth of cavity, $N_p$ the number of laser scans and $d_t$ the average depth obtained from the 1D trench profile using same milling parameters as described in section 3.2.

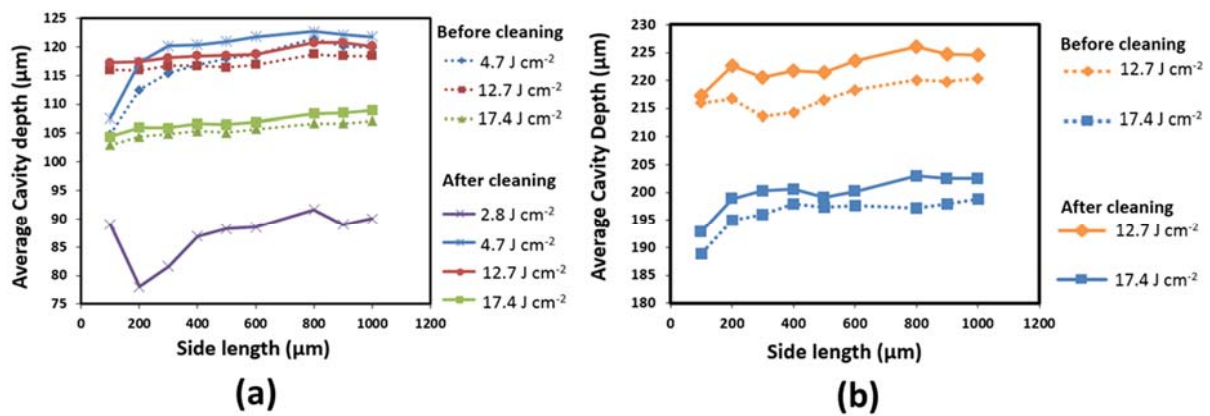

**Fig. 9.** Average depth of cavities milled on the 1030 nm source at repetition rate of 30 kHz for different fluences before and after ultrasonic cleaning (a) Cavities with depth ~100 µm (b) Cavities with depth ~ 200 µm

**Table 4.** Calculated quantities for milled cavities (multiple scans) on 1030 nm source at repetition rate of 30 kHz and scan speed of 20 mm s$^{-1}$

| Fluence (J cm$^{-2}$) | No of scans | Depth (µm) | Average roughness (µm) | RMS roughness (µm) | Efficiency factor | VRR* (x10$^6$ µm$^3$ s$^{-1}$) |
|---|---|---|---|---|---|---|
| 2.8 | 65 | 90 | 9.9 | 11.3 | 0.68 | 0.26 |
| 4.7 | 28 | 121.8 | 3.9 | 5.1 | 0.94 | 0.83 |
| 12.7 | 17 | 120.1 | 4.3 | 5.5 | 0.91 | 1.35 |
| 12.7 | 34 | 224.5 | 5.1 | 6.4 | 0.85 | 1.26 |
| 17.4 | 10 | 108.9 | 4.8 | 6.1 | 0.92 | 2.07 |
| 17.4 | 20 | 202.5 | 5.9 | 7.3 | 0.85 | 1.93 |

*VRR – Volume removal rate

The different parameters computed for 1mm square cavity are summarized in Table 4. The efficiency factors at fluences of 12.7 and 17.4 J cm$^{-2}$ are found to be 0.91 and 0.92, respectively, when the cavity depth is ~100 μm. On doubling the number of passes, the efficiency factor reduces to 0.85. Thus, it can be said that with increasing depth of the cavities, ablation plume and redeposited material play important role in determining ablated depth and roughness. With the reduction in efficiency factor, the surface roughness also increases. This is clearly noticed for the fluence of 2.8 J cm$^{-2}$, where the efficiency factor is much smaller than other cases. The volume removal rate (VRR) is increased as compared to that obtained single pass (in Table 3). This is because the first pass is on a pristine surface with good crystalline order and material removal is more difficult. In the subsequent passes, the structure has been already been significantly damaged or in other words there is incubation effect which increases the material removal rate [48]. The only exception again is fluence of 2.8 J cm$^{-2}$ because high volume of ablation plume in the form of debris impedes material removal. When the quantity of redeposited debris is small, the quality of side walls is good as seen in Fig. 10.

It has been shown in this section that choice of fluence is important in milling. Low fluence results in particle debris and surfaces which have high roughness. High fluence reduces particle debris to a great extent. However, with increasing fluence roughness also increases. Hence, it is beneficial to use the lowest possible fluence which does not give rise to particle debris. Also, when all other parameters are kept constant, the roughness increases with number of laser scans due to the build-up of ablation plume and redeposited material. Having discussed the different aspects of milling process, its application to SOI-CMOS technology is elaborated in the next section.

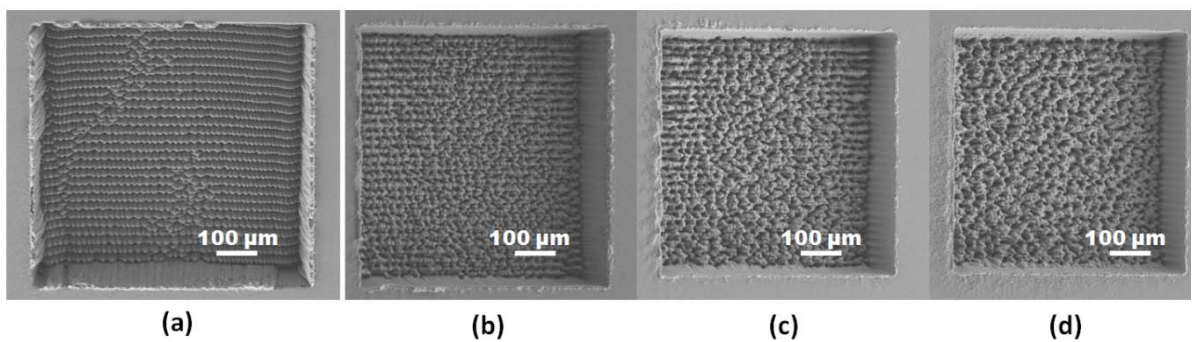

(a)   (b)   (c)   (d)

**Fig. 10.** SEM image of cavity side walls after ultrasonic cleaning. Milling parameters are: Scan speed - 20 mm s$^{-1}$, Repetition rate - 30 kHz, Fluence/ Number of scans (a) 2.8 J cm$^{-2}$/65 (b) 4.7 J cm$^{-2}$/28 (c) 12.7 J cm$^{-2}$/17 (d) 17.4 J cm$^{-2}$/10

## 5. SOI-CMOS Applications

The previous sections have dealt with the fabrication of cavities. For application in SOI-CMOS technology, an etch-protect layer is first laminated onto the handler side of the SOI die. Milling is then performed in the area where membranes need to be fabricated. In order to speed up the process, a two-step milling strategy has been used in the applications. In the first step, a cavity with larger area is created at a higher removal rate (fast step). In the second step, the smaller cavity of required dimensions of membrane is created at a slower removal rate (slow step). The bigger cavity allows more space for the diffusion of ablation plume generated during milling of smaller cavity which helps lower surface roughness. The die is then subject to XeF$_2$ etching to complete the removal of remaining few tens of microns of silicon remaining under smaller cavities and suspended membranes are obtained. The membranes of 3 different circuit components are studied: RF switch, RF isolation structures and Low-noise amplifier (LNA). The RF switch is a module in transceiver systems which contain large transistors for switching between transmission and reception of RF signals. RF isolation structures are specially designed structures to quantify RF coupling through the substrate. LNA is present in RF receivers to have a good overall noise performance of the receiver chain. A detailed treatment of these circuits can be found in [44,49].

Two different application demonstrations employing two-steps ablation are shown in Fig. 11. The milling parameters used for each case is shown in Table 5. The first case demonstrates possibility of milling close to the full thickness of SOI die (~750 µm). The mm sized cavity is shown in Fig. 11a is for the fabrication of membranes of RF switch circuit. For the RF switch, a single small cavity is milled within a large cavity and total depth of etching is 635 µm. The second case where two-step milling can be used for multiple membranes is shown in Fig. 11b applied to the study of RF isolation structures. Here, milling is performed on reduced starting thickness of the SOI die. The SOI die is grinded to have a thickness of 250 µm prior to the laser milling process. For RF isolation structures, four smaller cavities are milled within a large cavity and total depth of milling is 185 µm. It can be seen that by using a slower removal step within the larger cavity, roughness parameters are better for the smaller cavity as compared to larger one. Hence, a two-step strategy enables to obtain a good balance between removal rate and roughness.

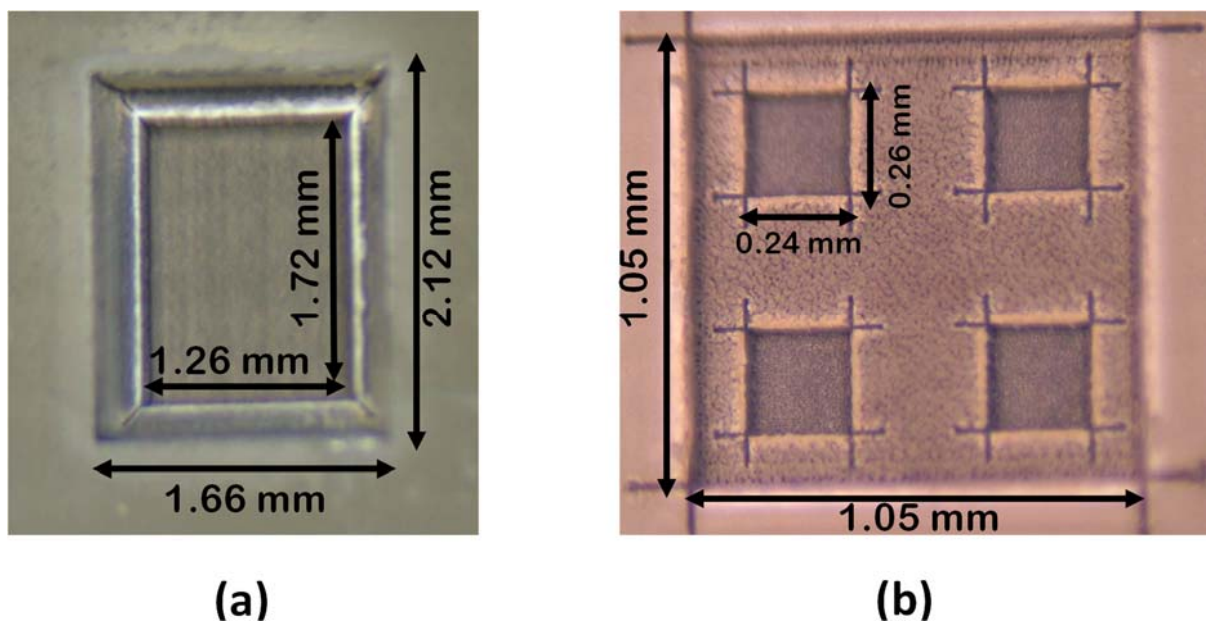

**Fig. 11.** Optical microscope images of two step milling for SOI-CMOS application for (a) RF switch and (b) Isolation structures

These two examples show that laser milling supports fabrication of membranes on varying die sizes and x-y dimensions. By using $XeF_2$ etching, the silicon inside the inner cavities can be completely etched and the BOX remains intact because it does not react with $XeF_2$. The etch-protect layer leaves the rest of the die intact. Additional details of $XeF_2$ etching are provided in [16]. After $XeF_2$ etch, free standing membranes of circuits are obtained. Since, the circuits are suspended on a transparent layer of BOX, light from the backside can be viewed on the front side. This allows to create an image of the suspended membrane of the RF functions. The membranes of different circuits are shown in Fig. 12. RF characterization of these circuits have been performed to quantify the performance improvement. Two types of SOI substrates are used for comparison: i) high resistivity SOI (HR-SOI) substrate and ii) trap rich SOI (TR-SOI) substrates. These are the most advanced substrates for RF applications on SOI-CMOS and are described in detail in [50].

**Table 5.** Different parameters in the 2-step fabrication of cavities on SOI wafer for switch and isolation structures

| Parameter | Switch | | Isolation structure | |
|---|---|---|---|---|
| | Fast step | Slow step | Fast step | Slow step |
| Scanner speed (mm s$^{-1}$) | 30 | 20 | 20 | 5 |
| Pulse repetition rate (kHz) | 25 | 40 | 30 | 2 |
| Laser power (W) | 1.45 | 0.26 | 1.006 | 0.013 |
| Number of passes | 10 | 30 | 4 | 180 |
| Fluence (J cm$^{-2}$) | 34.7 | 8.3 | 32.1 | 6.4 |
| Focus change per pass (μm) | 50 | - | - | 0.6 |
| Cavity Depth (μm) | 450 | 185 | 100 | 85 |
| RMS roughness (μm) | 15.5 | 12.8 | 8.3 | 6.1 |
| Removal rate (x 10$^6$ μm$^3$ s$^{-1}$) | 6.87 | 0.35 | 4.75 | 0.02 |

Many performance improvements are observed for different RF circuits as outlined in Table 6. For isolation structures, there is improvement in isolation (measured by ΔS$_{21}$). A reduction of S$_{21}$ by 7 dB and 5 dB at a frequency of 5 GHz for HR-SOI and TR-SOI respectively signifies that the stray coupling through the substrate is reduced after substrate removal. For RF switch, there is improvement in large signal linearity (2$^{nd}$ and 3$^{rd}$ harmonics) and also small signal losses (insertion and return loss) on both substrate types. The 2$^{nd}$ Harmonic improvement of ~23 dB and insertion loss improvement of 0.4 dB on HR-SOI substrate are the most significant results. Better performance of the switch can be attributed to elimination of substrate coupling in the OFF branches of the switch, which is the source of performance degradation. The Q-factor of inductors is improved significantly on membranes. A near doubling of Q-factor is observed for single turn inductor on HR-SOI substrate indicating amelioration of substrate losses. Such inductors with improved Q factor can help boost the performance of different RF circuits. For instance, inductor membranes have been used in LNA circuits for improvement of noise figure and linearity (IIP3). A 0.1 dB improvement in noise figure and 0.5 dB in IIP$_3$ as a result of improved quality factor of inductors successfully demonstrates the applicability of membranes in circuits of practical interest like LNA. Further details about electrical characterization of different circuits can be found in [44].

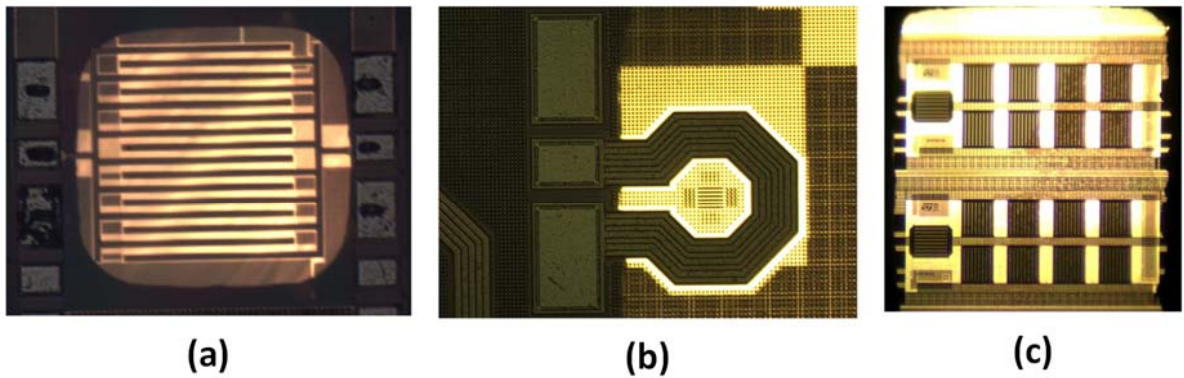

**Fig. 12.** Optical microscope pictures of membranes on SOI wafers obtained by passing backlight through the circuits (a) Isolation structure (b) Inductor (c) RF switch

Table 6. RF performance improvements obtained on membranes on two different types of SOI wafers

| Component | Parameter | Improvement on membrane | |
|---|---|---|---|
| | | HR-SOI | TR-SOI |
| *Isolation structures* | $\Delta S_{21}$ @ 5 GHz | 7 | 5 |
| *Switch* | 2nd Harmonic (dBm) | 22.7 | 5.6 |
| | 3rd Harmonic (dBm) | 7.8 | NI |
| | Return Loss (dB) | 4 | 4 |
| | Insertion Loss (dB) | 0.4 | 0.3 |
| *Inductors* | Q-factor (1 turn) (%) | 90 | 80 |
| | Q-factor (2 turns) (%) | 35 - 60 | 75 - 85 |
| *LNA (using membranes of inductors)* | Noise Figure (dB) | 0.1 | - |
| | IIP3 (dBm) | 0.5 | - |
| | $P_{1dB}$ (dBm) | NI | - |

NI – No improvement

# Conclusion

Large-area femtosecond laser milling of silicon has been reported which employs trench analysis. Five measures have been analysed for the trenches: i) average depth ii) recast layer volume (iii) average roughness (iv) trench width and (v) volume removal rate. Optical profilometry has been systematically used to obtain accurate and reliable measurements of all the parameters under study. Using these analyses, fluence and repetition rate conditions have been found suitable for laser milling. Further study on laser milling of cavities have shown additional factors to be taken into consideration. It has been found that presence of debris and material redeposition affects the quality of milling and suitable process parameters have been outlined. Application of large area milling has been demonstrated for RF circuits in SOI-CMOS technology. Suspended membranes of different circuits like RF switch, isolation structures, inductors and LNA have been successfully fabricated. The obtained performance is better than state-of-the-art SOI wafers for RF technology demonstrating the capability of laser processing for microelectronics applications.

# Acknowledgements

This work was supported by the French government through the National Research Agency (ANR) under program PIA EQUIPEX LEAF ANR-11-EQPX-0025, the STMicroelectronics-IEMN common laboratory and the French RENATECH network on micro and nanotechnologies. The authors also thank Oxford Lasers Ltd for helpful discussions on instrumental developments and laser processing techniques.

# APPENDIX

## A.1: Graphical illustration of measured trench parameters

The different measured parameters are depicted in Fig. A.1. The trench profile roughness is determined from the axial trench profile using the formula (6). The inner and outer trench width correspond to the measured excluding and including recast layer width respectively. The recast layer volume is the entire volume projecting over the substrate level and normalized to unit length of scribed trench.

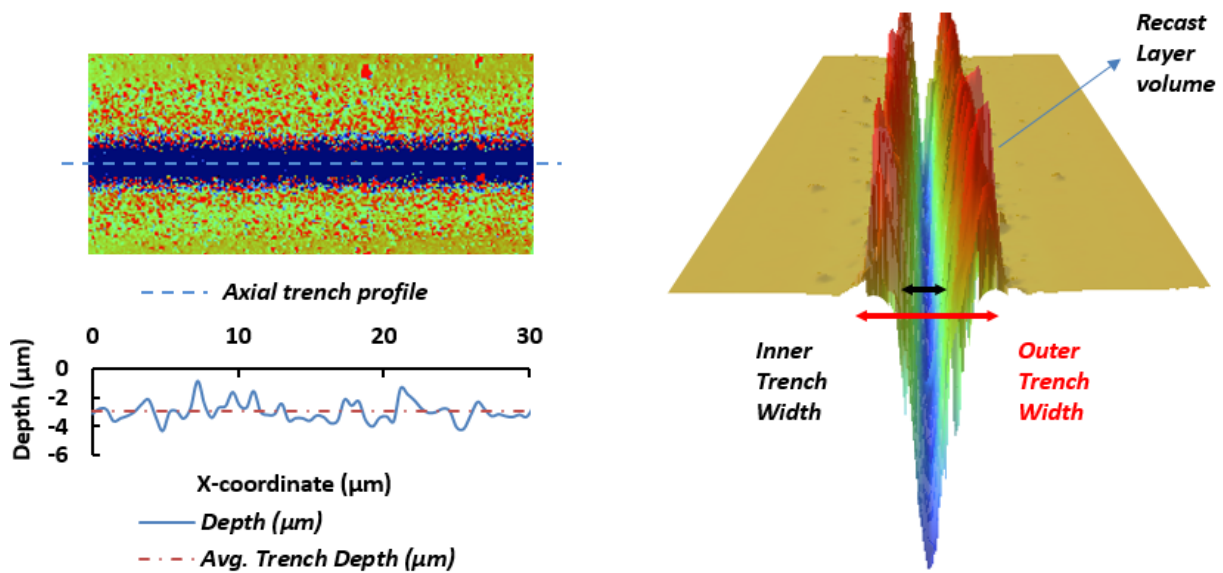

**Fig. A.1:** The trench profile shown as contour plot is shown on the left along with the axial trench profile from which average trench depth and the trench profile roughness parameters are deduced. On the right the trench widths and recast layer volume is shown on a 3D profile

## A.2: Trenches in the two regimes

The contour plots of trench profile for the two regimes shown in Fig. A.2. The red points correspond to the recast layer. A clear and prominent recast layer can be observed for the strong ablation regime as compared to gentle ablation regime.

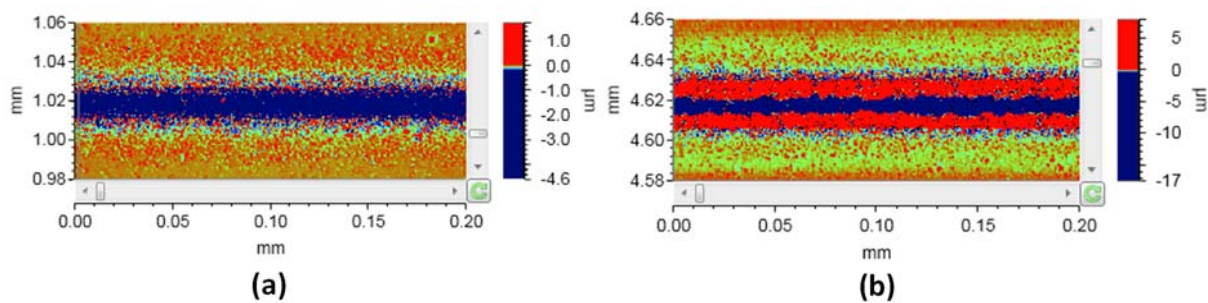

**Fig. A.2:** Contour plots of trench profile scribed on 1030 nm source at a repetition rate of 200 kHz and scan speed of 50 mm s$^{-1}$ (a) Gentle ablation regime (Fluence = 1.27 J cm$^{-2}$) (b) Strong ablation regime (Fluence = 7.07 J cm$^{-2}$)

## A.3: Debris along the trench axis

The profile taken for a 200 kHz repetition rate and 1030 nm source at a scan speed of 50 mm s$^{-1}$ and fluence of 12.1 J cm$^{-2}$ is shown in Fig. A.1. The axial trench profile shows an indication of presence of debris. There are locally deep regions of depth ~25 µm. This is likely the depth at which there is penetration of laser energy and ablation occurs. However, the presence of debris along the axis of the trench can be seen in the form of lower local depths and protrusions above the trench. This causes the measured average depth to be much smaller (~5 µm). It can be also seen that the trench opening is really narrow because of a prominent recast layer.

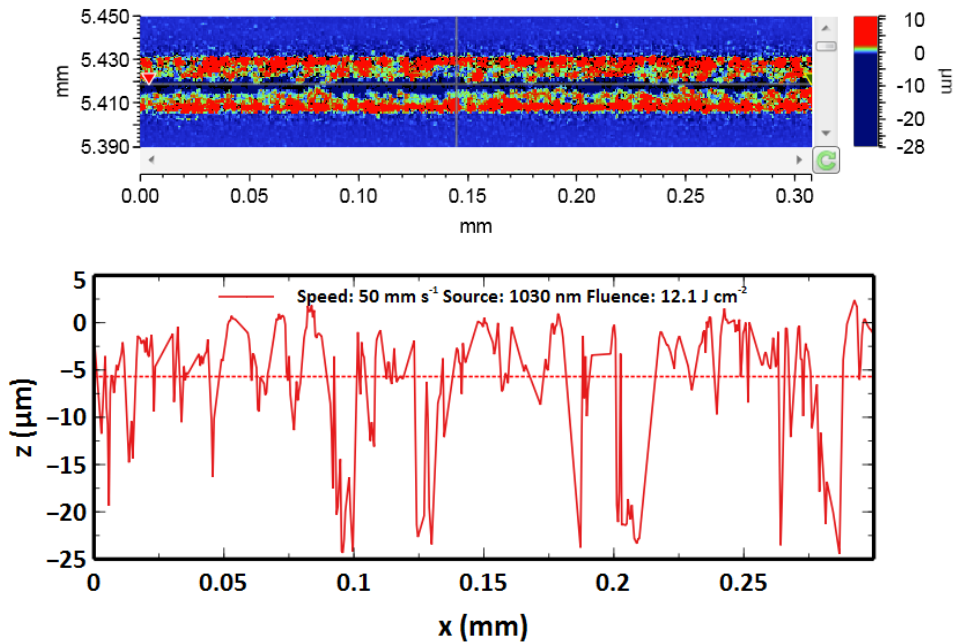

**Fig. A.3:** Axial trench profile showing narrowing of trench and possible presence of debris along trench axis

## A.4: SEM scans of trenches showing periodic profiles

The SEM scans for the three cases studied in trench scribing experiment is shown in Fig. A.2. The experimental parameters chosen for trenches shown in Fig. A.2 results in an average depth of ~5 µm. Etched profiles with a periodic distribution of holes are clearly visible for both 343 nm and 1030 nm source at a repetition rate of 200 kHz. It can also be seen that with increase in scan speed, the period increases. For the 1030 nm source at a repetition rate of 30 kHz, no distinct periodic hole pattern is observed.

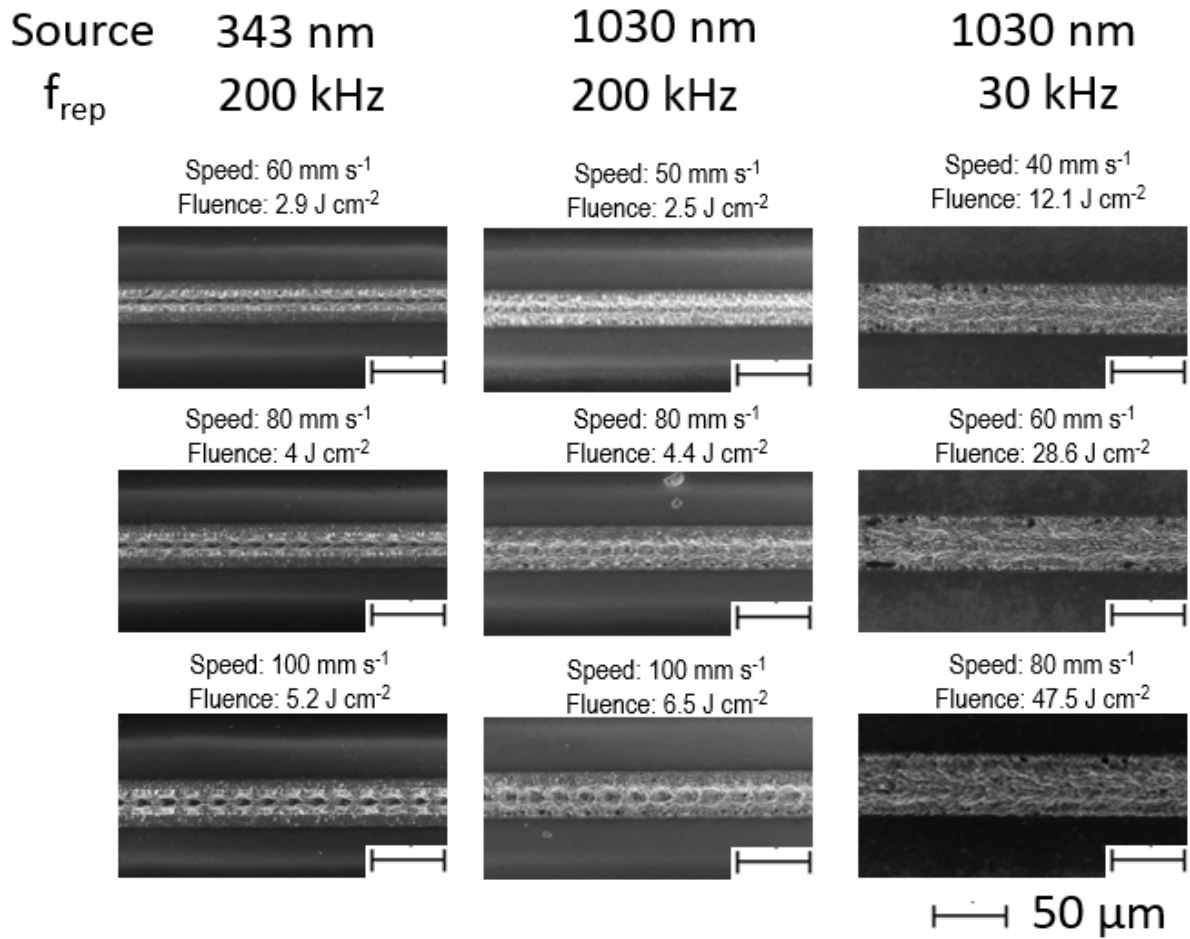

**Fig. A.4:** SEM scans of trenches of average depth ~5 μm with different process parameters